\documentclass[showkeys,showpacs,nosuperscriptaddress,twocolumn,floatfix]{revtex4}
\usepackage{graphicx}
\usepackage{psfrag}
\usepackage{amsmath}
\usepackage{amsfonts}
\usepackage{color}
\usepackage{hyperref}
\usepackage{listings}

\begin{document}

\title{Zaller--Deffuant model of mass opinion}

\date{\today} 

\author{Krzysztof Malarz}
\homepage{http://home.agh.edu.pl/malarz/}
\email{malarz@agh.edu.pl}

\author{Piotr Gronek}
\email{gronek@agh.edu.pl} 
 
\author{Krzysztof Ku{\l}akowski}
\email{kulakowski@novell.ftj.agh.edu.pl}

\affiliation{
Faculty of Physics and Applied Computer Science,
AGH University of Science and Technology,\\
al. Mickiewicza 30, PL-30059 Krak\'ow, Poland
}

\date{\today}
 
\begin{abstract}
Recent formulation of the Zaller model of mass opinion is generalized to include the interaction between agents. 
The mechanism of interaction is close to the bounded confidence model. The outcome of the simulation is the probability distribution of opinions on a given issue as dependent on the mental capacity of agents. Former result was that a small capacity leads to a strong belief. Here we show that an intensive interaction between agents also leads to a consensus, accepted without doubts. 
\end{abstract}
 
 

\maketitle
  
\section{Introduction}
Mathematical modeling of the dynamics of public opinion \cite{sta,rev}
becomes an autonomous area in computational social sciences \cite{b1,b2,b3}. Most established \cite{rev} are the voter model \cite{voter}, the Galam model \cite{gal}, the social impact model \cite{lev}, the Sznajd model \cite{sznajd}, the Deffuant model \cite{def} and the Krause--Hegselman model \cite{kh}. In these models, opinions are represented by numbers, either integer \cite{voter,gal,lev,sznajd} or real \cite{def,kh}. Recently a new perspective was proposed by Martins \cite{mar}: an agent is represented by a continuous distribution of opinions, and his binary decisions are 
formulated on the basis of this internal information. Basically, the issue of the modeling is a consensus between agents, and their decisions are taken on the basis on the information available in the system at the beginning. In all models constructed on the basis of statistical physics \cite{rev}, opinions are exchanged between agents. However, in real situations the decisions are influenced by information
flowing continuously from mass media, and the way how the informations are produced, selected and shaped largely determines the way the audience understand the world \cite{lipp,mcl,aro,chom,zal,lis}. This role of mass media cannot be overestimated; the term ``global village'' was coined by Marshall McLuhan as early as in 60's, and today it is even more appropriate. On the contrary to models inspired by physics, in social sciences theory of the public opinion is concentrated on media. Such is also the Zaller model, known also as the Receipt-Accept-Sample (RAS) model \cite{zal}; this is perhaps the most influential mathematical model of the public opinion.

The Zaller model \cite{zal} is an attempt to describe the processes of receipt messages from media by the audience, 
of accepting these messages or not, and to use them as a basis to formulate binary (Yes or No) decisions. In its original 
formulation \cite{zal} the model is a parameterization; based on poll results, a set of phenomenological coefficients can be 
assigned to each issue. More general knowledge could be attained by observing some regularities in these coefficients.
As a mathematical project, the program demands a large scale investigation of statistical data, but it offers little insight for a sociophysical research, which is oriented towards mechanisms. Still, in its descriptive content the Zaller model is an invaluable starting point to advance our understanding of the processes listed above. The approach is rightly termed as the
Receipt-Accept-Sample (RAS) model.

Recently the model was reformulated to a more geometrical form \cite{ja}, which
applies the effect of bounded confidence \cite{def} to the stream of messages coming from media \cite{zal}. In this new formulation, information accumulated by agents is encoded in the form of a probability distribution; this is similar to the approach by Martins \cite{mar}. The motivation was to free the construction from tens of parameters, which in the original formulation \cite{zal} are to be obtained by fitting the model curves to poll data. This target was reached by introducing a plane of issues, where particular opinions and messages were represented by points on the plane. The distance between points A and B on the plane was a measure of the difficulty of receipt an opinion A for an agent with an opinion B. This construction was adopted from the Deffuant model \cite{def} and it is known as the bounded confidence: if the distance between two agents is larger than some threshold value, these agents ignore each other. That is why here we use the term ``Zaller--Deffuant model''. In both formulations \cite{zal,ja}, the contact between agents is substituted by a stream of messages, provided by media. Still, the interaction between agents---the mechanism so basic for all sociophysical models---is absent in both formulations \cite{zal,ja}.

The aim of this paper is to generalize the model \cite{ja} by adding an interaction between agents. This is built-in, retaining the bounded confidence effect. As in \cite{ja}, here  the outcome of the calculation is the probability distribution of the external decision, Yes or No. The only parameter is the mental capacity $\mu$ of agents, which is a measure of a maximal distance from messages received previously to a newly received message. Its detailed definition is in the next section.  In the Deffuant model \cite{def}, the similarly defined parameter is the threshold.

In the next section we describe the geometrical form \cite{ja} of the model. The interaction between agents is introduced in two alternative versions: {\it i)} the interaction happens with some probability and the message is directed to all agents, {\it ii)} agents are distributed in nodes of a homogeneous random network and the interaction is directed only to neigbors. Section \ref{sec-3} is devoted to numerical results of both versions of the model. Discussion of the results closes the text.

\section{The model}
\begin{figure}
\includegraphics[width=0.48\textwidth,viewport=50 0 310 252,clip]{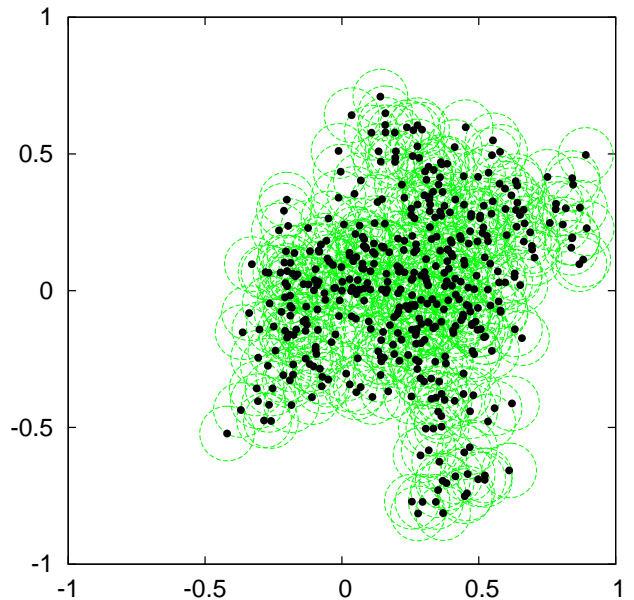}
\caption{The square $2\times 2$ on the plane of issues.
Set of messages received by a model agent with $\mu$=0.1 till a given time $t$.
With the simulation continued, the square will be sooner or later fulfilled; sooner for agents with larger $\mu$.}
\label{fig-1}
\end{figure}

Messages are represented as points on a plane. This means that the message is characterized by its relation to two issues, say economic and moral. The probability density function of positions of incoming messages is constant within a given area, say a square $2\times 2$, and it is zero outside the square.  Now consider a new message appears. Each agent has to decide: to receive the message or not. Once the position of the new message is too far to what the agent received in the past, the message is simply ignored. The critical distance between messages is the agent's mental capacity $\mu$; small value of this parameter means that the agent will be able to receive only messages found in the direct neighborhood of the messages received by him in the past. To express $\mu$ mathematically, let us consider an agent $i$ who received a series of $n_i$ messages before time $t$. Let us denote the coordinates of these messages on the plane of issues as $(x_i(t'),y_i(t'))$, where $t' < t$. Analogously, the coordinates of the new message are $(x(t),y(t))$. Then, the capacity $\mu$ is defined by the following relation:  $i-$th agent receives the new message if and only if $t''<t$ exists such that
\begin{equation} 
[x_i(t'')-x(t)]^2+[y_i(t'')-y(t)]^2<\mu^2.
\end{equation} 
In this way, at each time $t$ each agent $i$ is represented by 
the spatial distribution of messages $\rho_i(t)$ which he received till $t$: $\rho_i(x,y)=const \ne 0$ if the distance from $(x,y)$ to 
at least one message received previously by $i$ is smaller than $\mu$. A numerical example of messages received by an agent is shown in Fig. \ref{fig-1}. 
Time $t$ is conveniently equivalent to the number of incoming messages, not necessarily received.

\begin{figure}[ht]
\includegraphics[width=0.48\textwidth,viewport=50 0 310 252,clip]{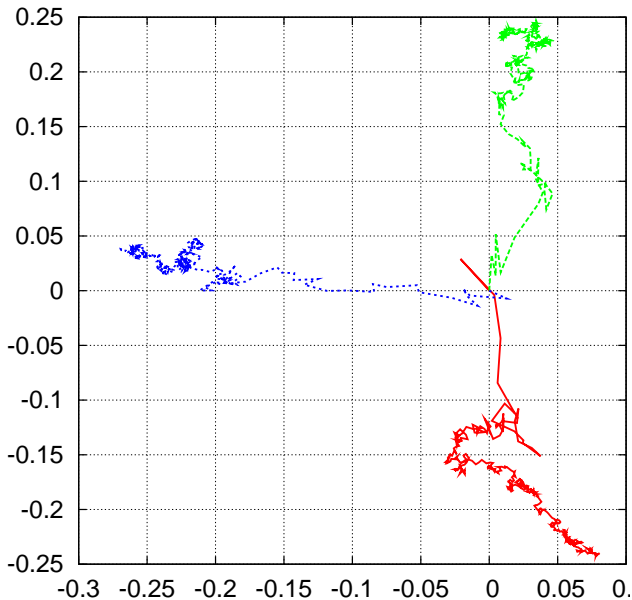}
\caption{Three examples of paths of mean opinions of agents with $\mu=0.1$, who strongly believe into a model issue; most of received messages are 
concentrated on the same side of the horizontal axis (OX) or the vertical axis (OY).}
\label{fig-2}
\end{figure}
 
The source codes of all variants of the calculations are available at \cite{naszastrona}.

\begin{figure*}[htb]
\includegraphics[width=0.45\textwidth,viewport=50 0 310 252,clip]{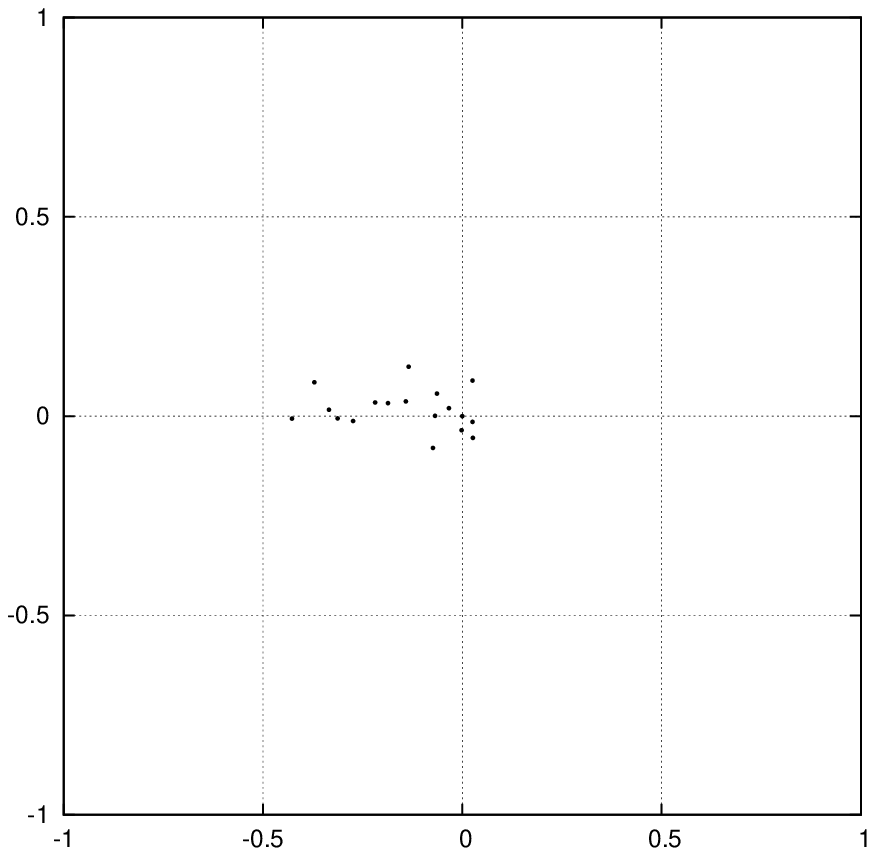}
\includegraphics[width=0.45\textwidth,viewport=50 0 310 252,clip]{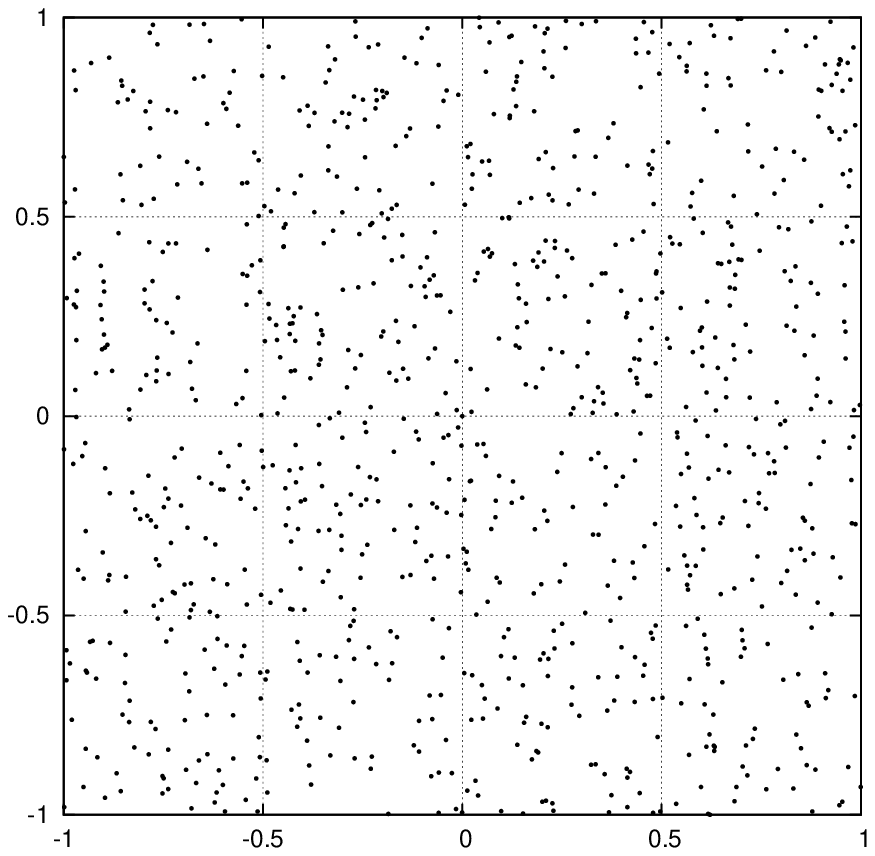}\\
\includegraphics[width=0.45\textwidth,viewport=50 0 310 252,clip]{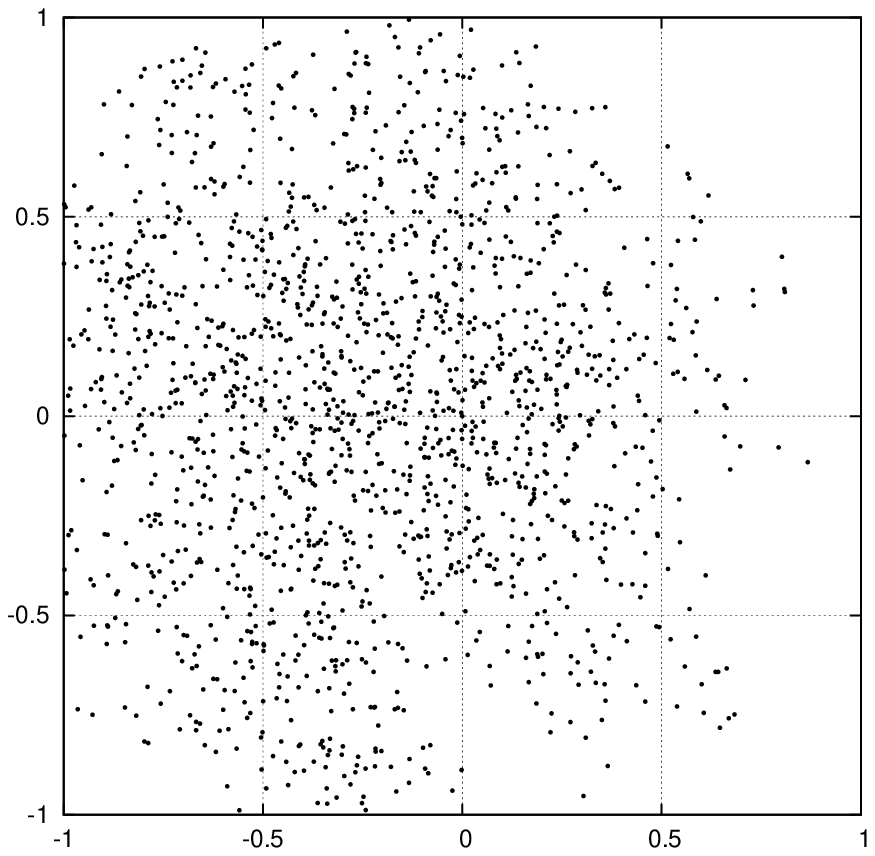}
\includegraphics[width=0.45\textwidth,viewport=50 0 310 252,clip]{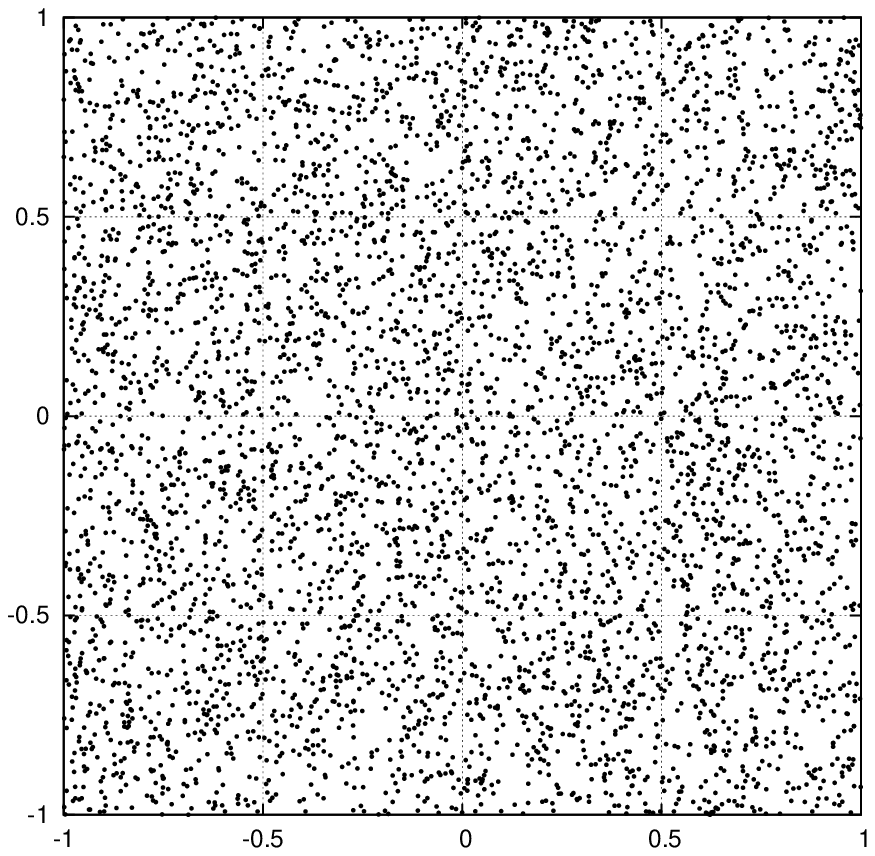}\\
\includegraphics[width=0.45\textwidth,viewport=50 0 310 252,clip]{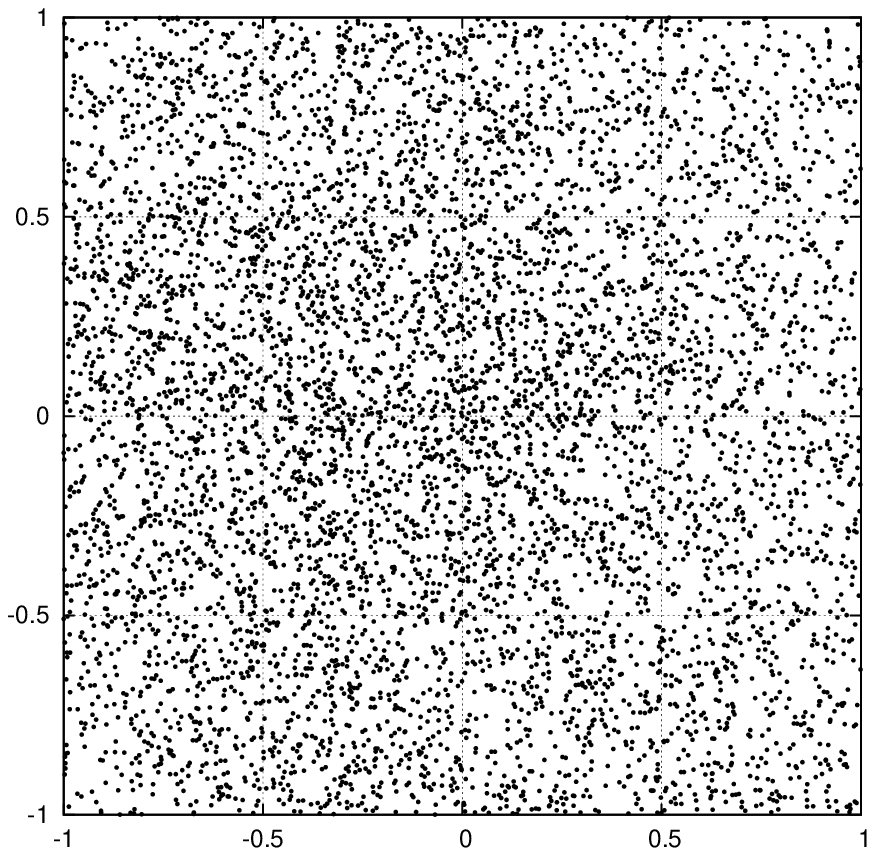}
\includegraphics[width=0.45\textwidth,viewport=50 0 310 252,clip]{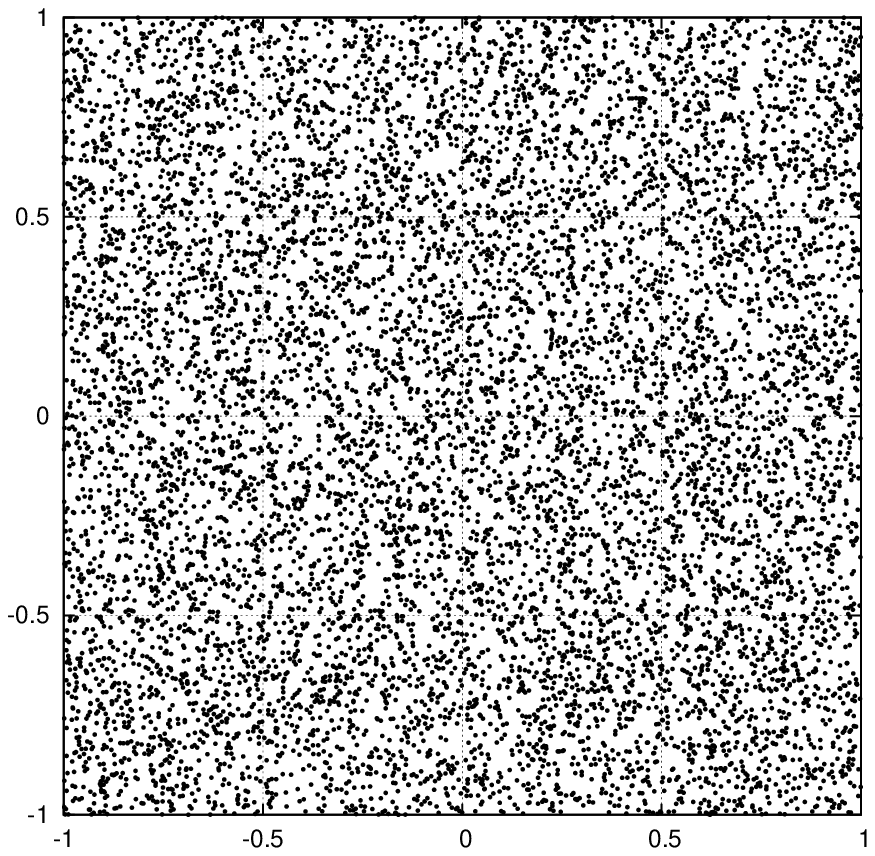}
\caption{\label{fig-messages-bez}Evolution of single agent knowledge.
Initially, agent is situated at the center of the square.
Subsequent rows correspond to sending $M=10^3$, $5000$ and $10^4$ messages.
Left column corresponds to $\mu=0.1$ while right is for parameter $\mu$ equal to 0.5.}
\end{figure*}

\begin{figure*}[htb]
\includegraphics[width=0.45\textwidth,viewport=50 0 310 252,clip]{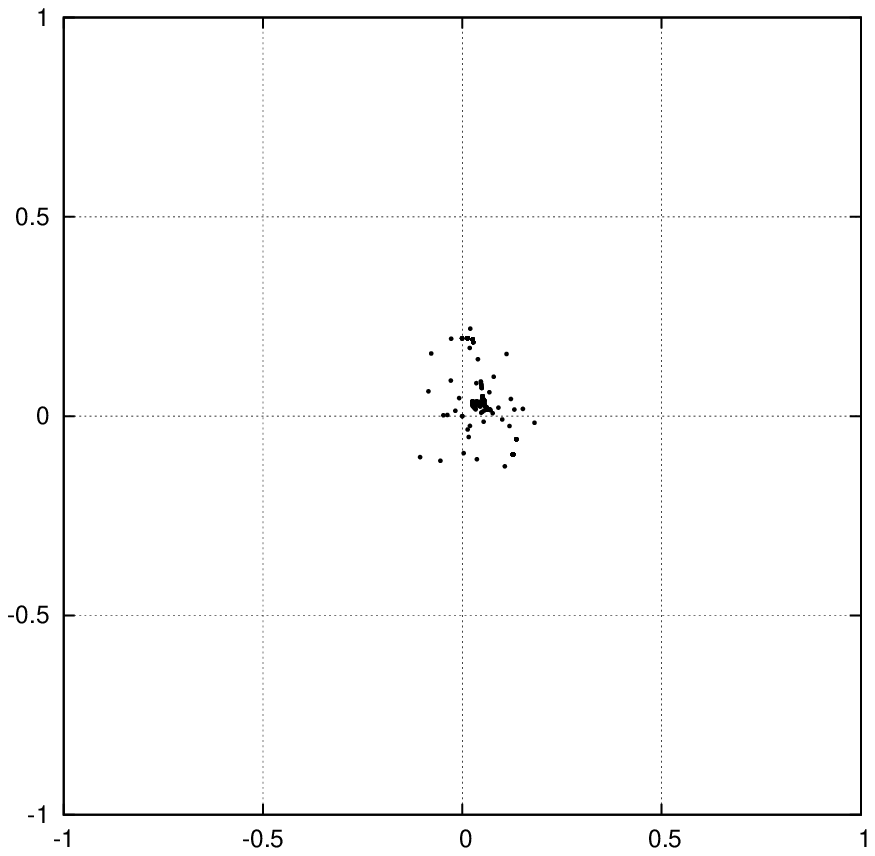}
\includegraphics[width=0.45\textwidth,viewport=50 0 310 252,clip]{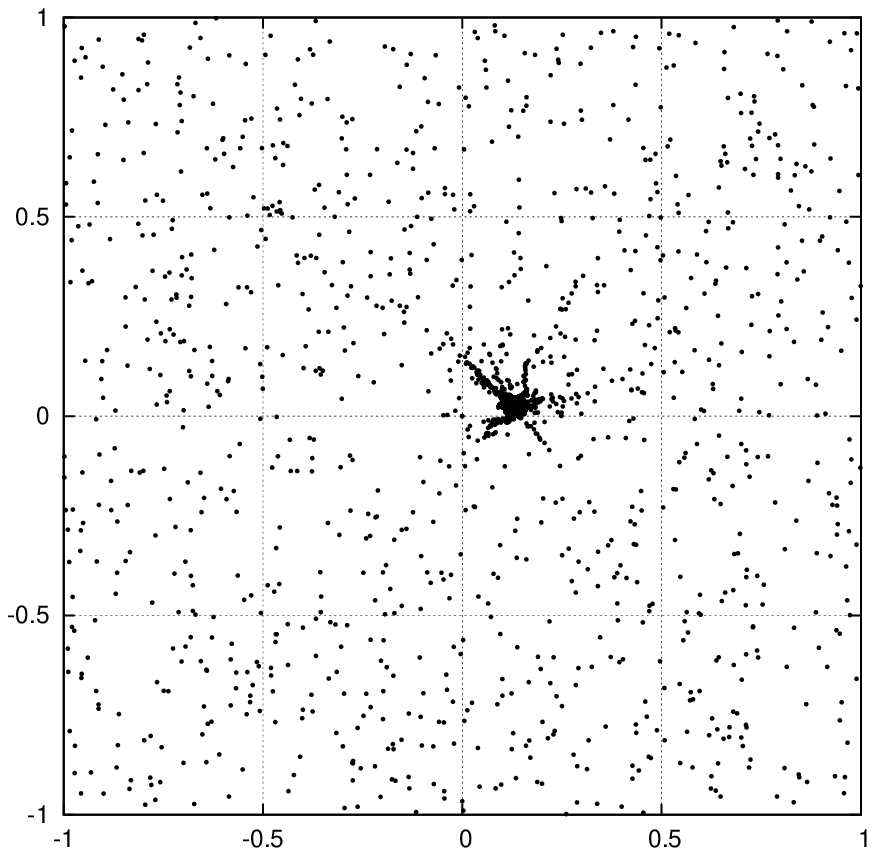}\\
\includegraphics[width=0.45\textwidth,viewport=50 0 310 252,clip]{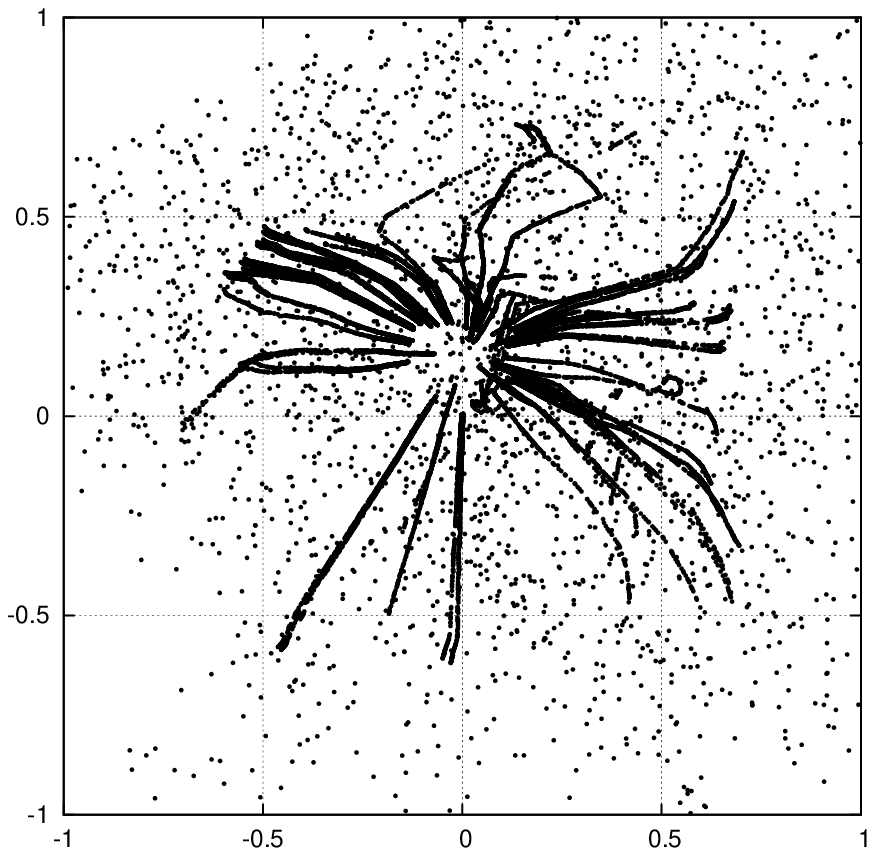}
\includegraphics[width=0.45\textwidth,viewport=50 0 310 252,clip]{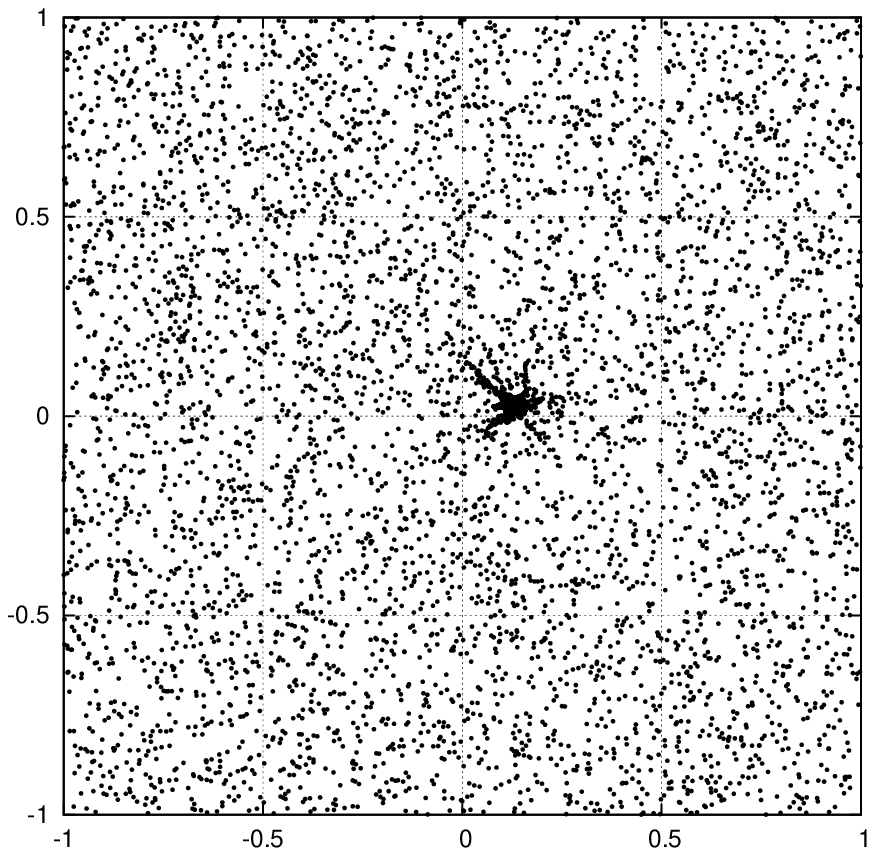}\\
\includegraphics[width=0.45\textwidth,viewport=50 0 310 252,clip]{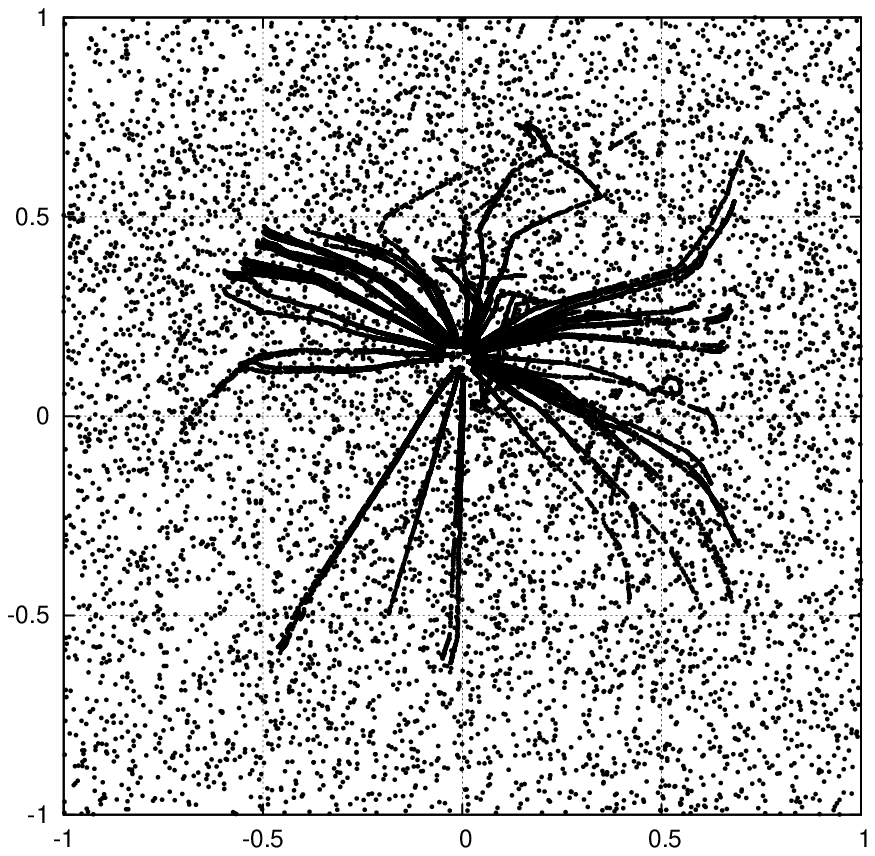}
\includegraphics[width=0.45\textwidth,viewport=50 0 310 252,clip]{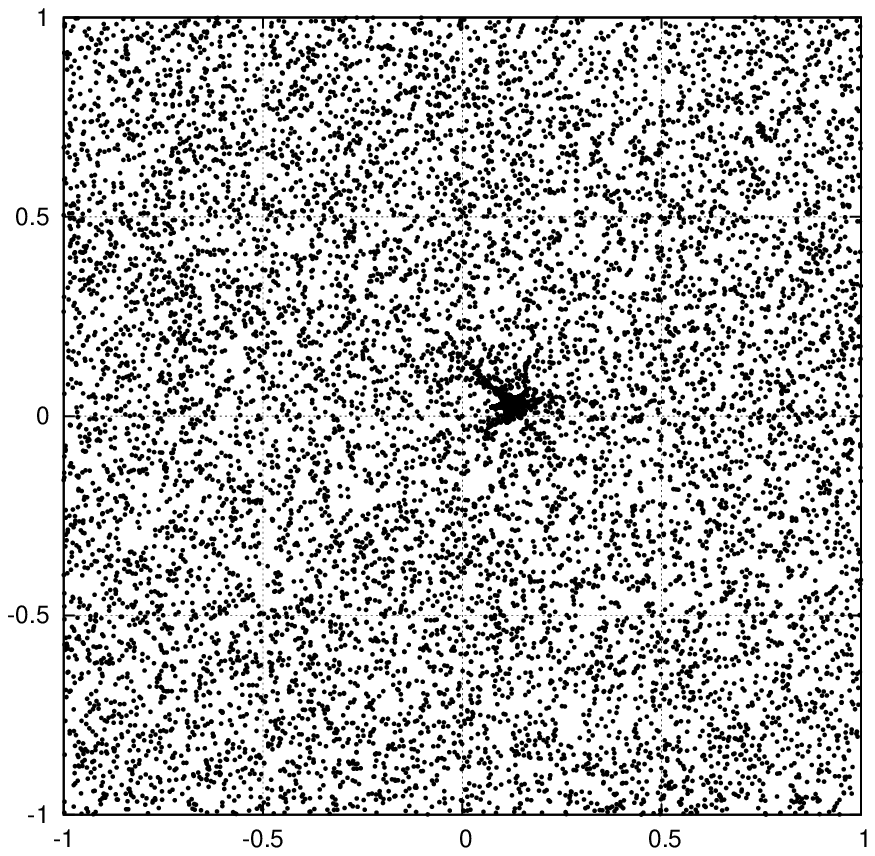}
\caption{\label{fig-messages-z}Evolution of single agent knowledge.
Initially, agent is situated at the center of the square.
Subsequent rows correspond to sending $M=10^3$, $5000$ and $10^4$ messages.
Left column corresponds to $\mu=0.1$ while right is for parameter $\mu$ equal to 0.5.
The agent interacts with 99 other agents.}
\end{figure*}

One of immediate consequences of this geometrical representation is that the probability of receipt a new message increases with the number of those already received. This is due to the fact that a new message is received if it is not too far from the area occupied by the messages previously received. As it is seen in Fig. \ref{fig-1}, this area increases with the number of received messages. On the other hand, the effect is in accordance with the second postulate of the Zaller theory \cite{zal}: {\it ...people are able to react critically to the arguments they encounter only to the extent that they are knowledgeable about political affairs.} They are knowledgeable---it means that they did receive some messages already in the past.

The idea to represent messages by points on a plane is not new and it was evoked at different occasions \cite{ksz,comp,sylv,dpra}. Consider an issue, about which the agent is asked by a pollster to construct an opinion. To do this, he makes a projection of the new issue to the two issues salient for him, which span his personal plane. This is equivalent to a new axis on the plane. All messages accepted by the agent can be projected on this new axis, and their projected density gives the probability of the answer Yes or No. In particular, if the new axis is chosen to be just the vertical one (OY), the normalized probability $p_i$ of answer Yes obtained from agent $i$ is equal to
\begin{equation}
p_i=\frac{\sum_jx_j(i)\Theta(x_j(i))}{\sum_j|x_j(i)|},
\end{equation}
where $x_j(i)$ is the $x$-th coordinate of the $j$-th message received by $i$-th agent, and $\Theta(x)$ is the step function defined as follows: $\Theta(x)=1$ for $x\ge 0$, otherwise $\Theta(x)=0$.

Our numerical experiment is to expose all agents to a homogeneous stream of messages, covering the area with equal density.
In the average, there is no more arguments for one option than for another. With this kind of information, a reasonable agent should stay undecided, what is equivalent to the decision Yes or No with the same probability. Let us consider an agent with capacity $\mu$ as large that the circle with radius $\mu$ covers the whole square where new messages appear, either from other agents or from media. Such an agent receives all messages; then after a short time $t$ he is represented by a function $\rho$ equally distributed on the square.  
Indeed, the probability $p$ of his/her Yes is close to 0.5. Now let us consider another agent with small $\mu$. The number of messages received by him increases only slowly, and so does the probability that he will receive a new message, measured by an area where his/her $\rho_i \ne 0$. The spatial distribution of his $\rho$ remains nonhomogeneous for a long time. Once a new axis of an issue is set, the projection of the received messages on this axis remains either mostly at positive or mostly at negative side. In combinatorics, the problem is
known as the first arcsine law \cite{fel}. As a result, either the answer Yes will be given with large probability and the answer No---with small probability, or the opposite. In other words, opinions of the agent with small $\mu$ are well established or extreme, what is not justified by the content of incoming messages. This is the result obtained numerically in \cite{ja}. The stage of opinion formation corresponds to the stages Accept-Sample in the RAS model \cite{zal}.

Now we are going to expand the model by adding the interaction between agents. This is done in two ways. In first version, after incoming of each message from media each agent sends his own message to everybody with probability $r$. The position of this message in a plane is equal to the average position of the informations received by the sender. The rule to receive this message by other agents or not is the same as before. In Fig. \ref{fig-2} we show three examples of how these average positions depend on the number of received messages, if the capacity $\mu$ is small. The message is handled by all other agents in the same way as the messages from media. In the second version, agents are placed at nodes of an Erd{\H o}s--R\'enyi network \cite{bb}. After incoming each external message, every agent sends the information on his average opinion as before, with probability $r=1$. The difference is that these messages can be received only by the sender's neighbours. In both versions, initial positions of the agents are randomly distributed, what reflects the commonly known spread of opinions. This is also an advantage with respect to our previous approach \cite{ja}, where each agent started from the centre of the square.

\section{Results \label{sec-3}}
\begin{figure}[ht]
\psfrag{mu}{$\mu=$}
\psfrag{p}{$p_i$}
\psfrag{P(p)}{$P(p_i)$}
\includegraphics[width=0.48\textwidth]{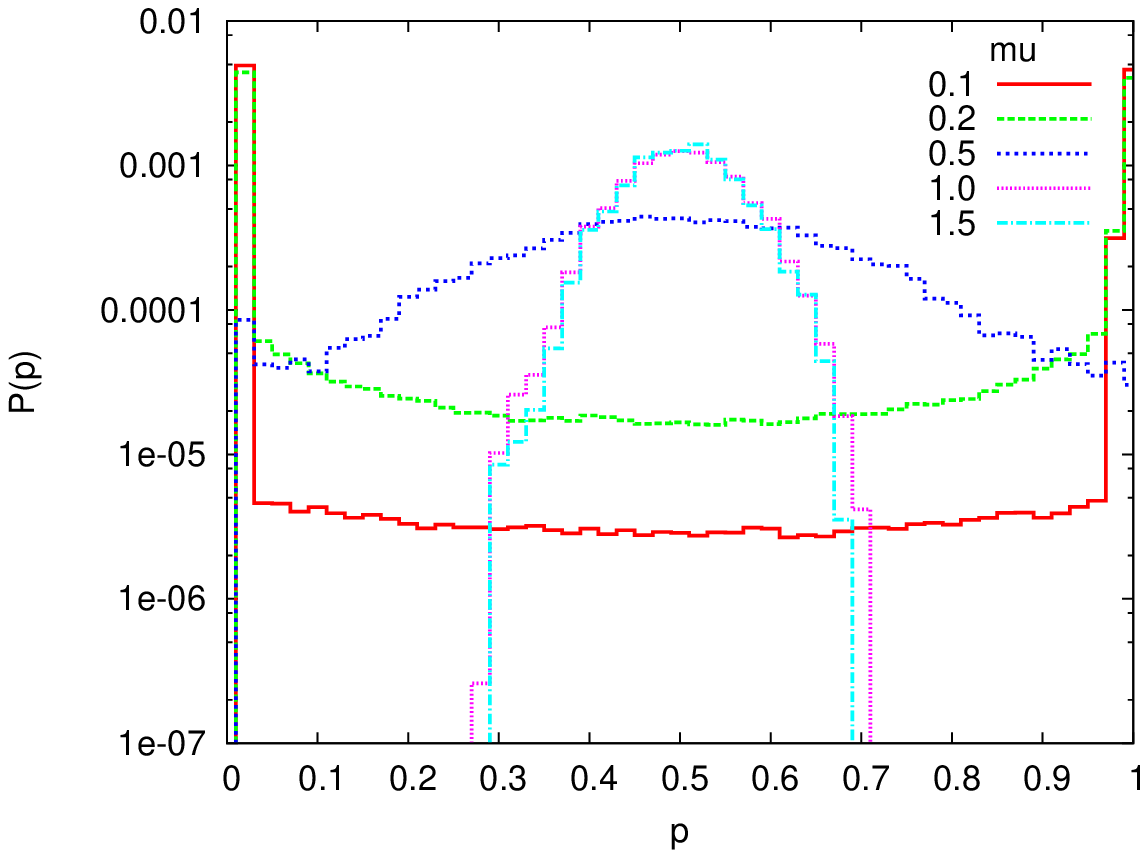}
\caption{The probability distribution of $p_i$ for various capacities $\mu$ without interactions between agents, for the probabilistic variant of the calculation.}
\label{fig-3}
\end{figure}

\begin{figure}[ht]
\psfrag{mu}{$\mu=$}
\psfrag{p}{$p_i$}
\psfrag{P(p)}{$P(p_i)$}
\includegraphics[width=0.48\textwidth]{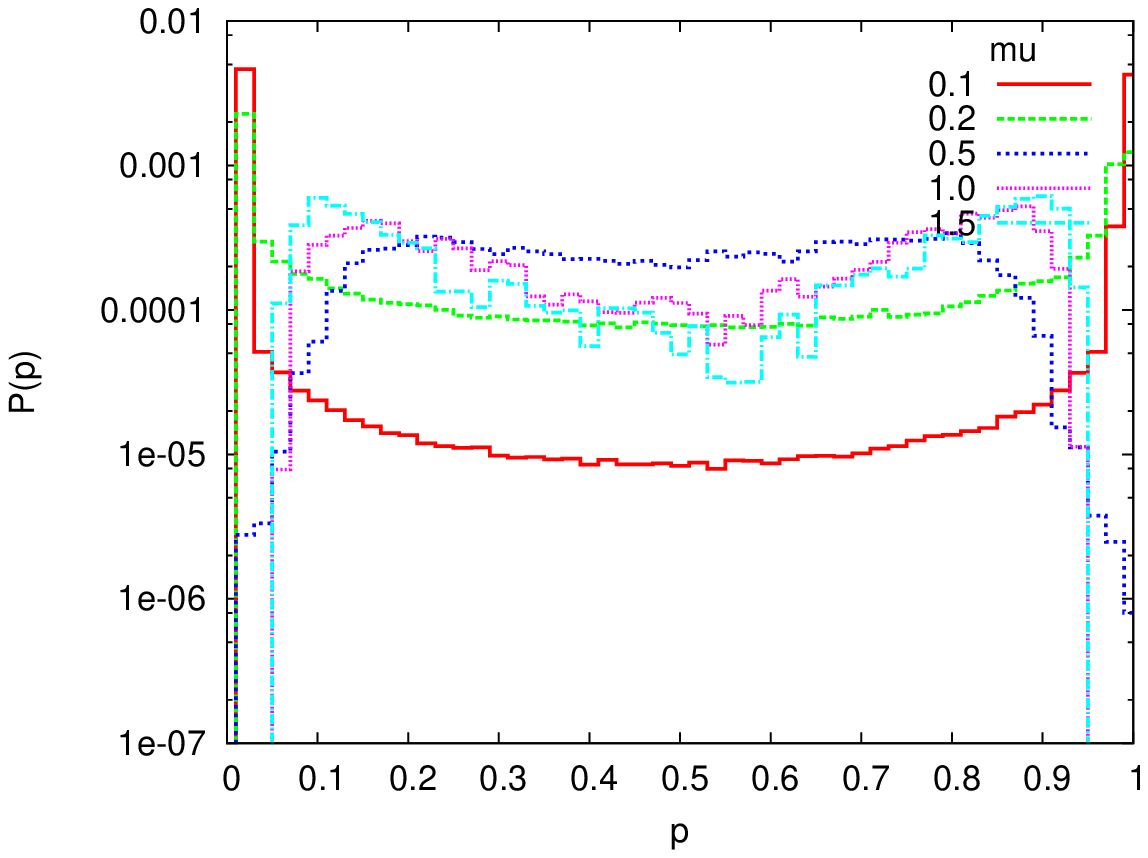}
\caption{The probability distribution of $p_i$ for various capacities $\mu$ with interactions between agents, for the probabilistic variant of the calculation.}
\label{fig-4}
\end{figure}

In Figs. \ref{fig-3} and \ref{fig-4} we show the results obtained for the probabilistic variant, where agents express their opinions with probability $r$. The number of agents is $10^3$, the number of external messages is $100$, and the results are averaged over $1000$ simulations. The value $r=10^{-2}$ is set to assure that about ten agents provide their messages per each message from the media.

We see that for small $\mu$ the interaction has no visible effect;  in both cases the histograms show strong maxima for $p_i\approx 0$ and $p_i\approx 1$. This means, that for small capacity, the messages produced by the agents are too far to be received. As a consequence, the opinions remain extreme: no doubts, Yes or No with absolute certainty. For large capacity $\mu$ the cases without and with the interaction are remarkably different. In the former case, a large maximum is visible at $p_i\approx 0.5$ for $\mu=1.0$ and $\mu=1.5$ (Fig. \ref{fig-3}). In the latter case (Fig. \ref{fig-4}), this maximum disappears. This result indicates, that in the case of an intensive interaction even the large capacity $\mu$ does not prevent opinions from being well established; they are just less extreme, than for small $\mu$.

\begin{figure}[ht]
\psfrag{mu}{$\mu=$}
\psfrag{p}{$p_i$}
\psfrag{P(p)}{$P(p_i)$}
\includegraphics[width=0.48\textwidth]{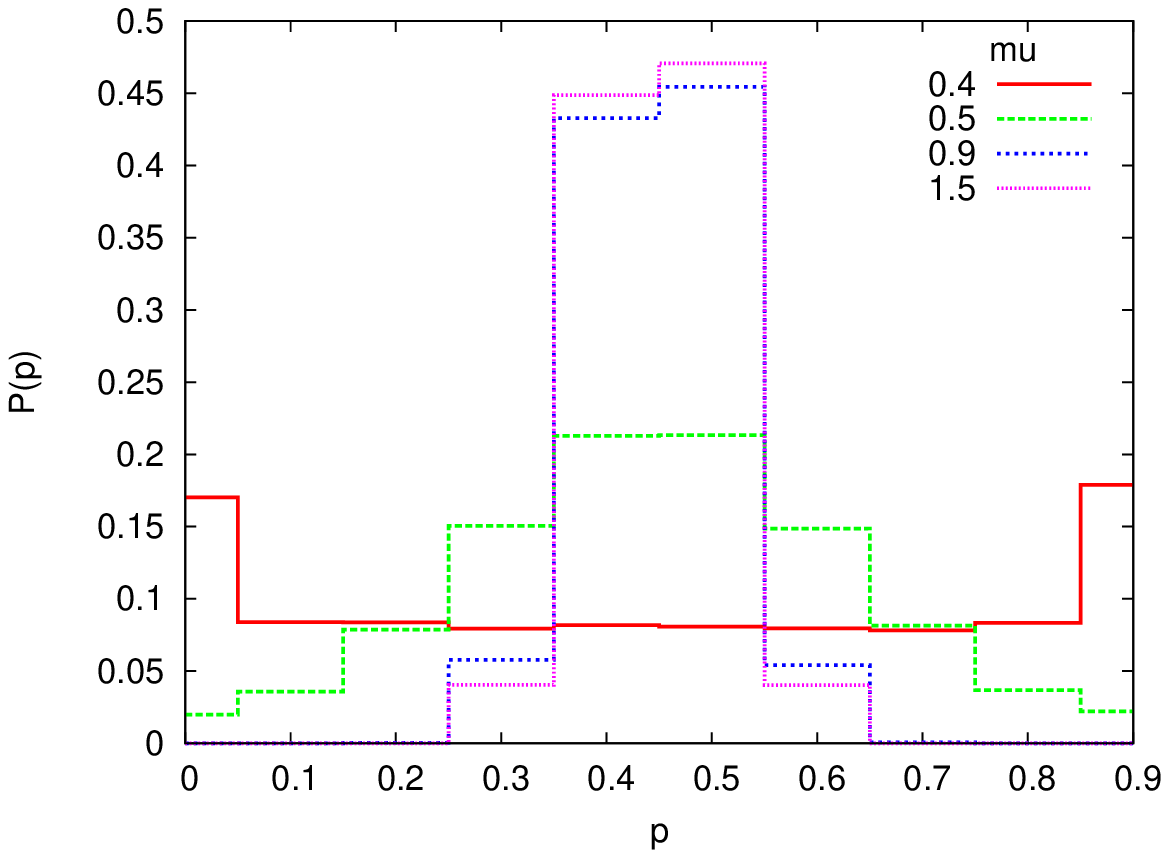}
\caption{The probability distribution of $p_i$ for various capacities $\mu$ without interaction between agents, distributed in a random network.}
\label{fig-5}
\end{figure}

\begin{figure}[ht]
\psfrag{mu}{$\mu=$}
\psfrag{p}{$p_i$}
\psfrag{P(p)}{$P(p_i)$}
\includegraphics[width=0.48\textwidth]{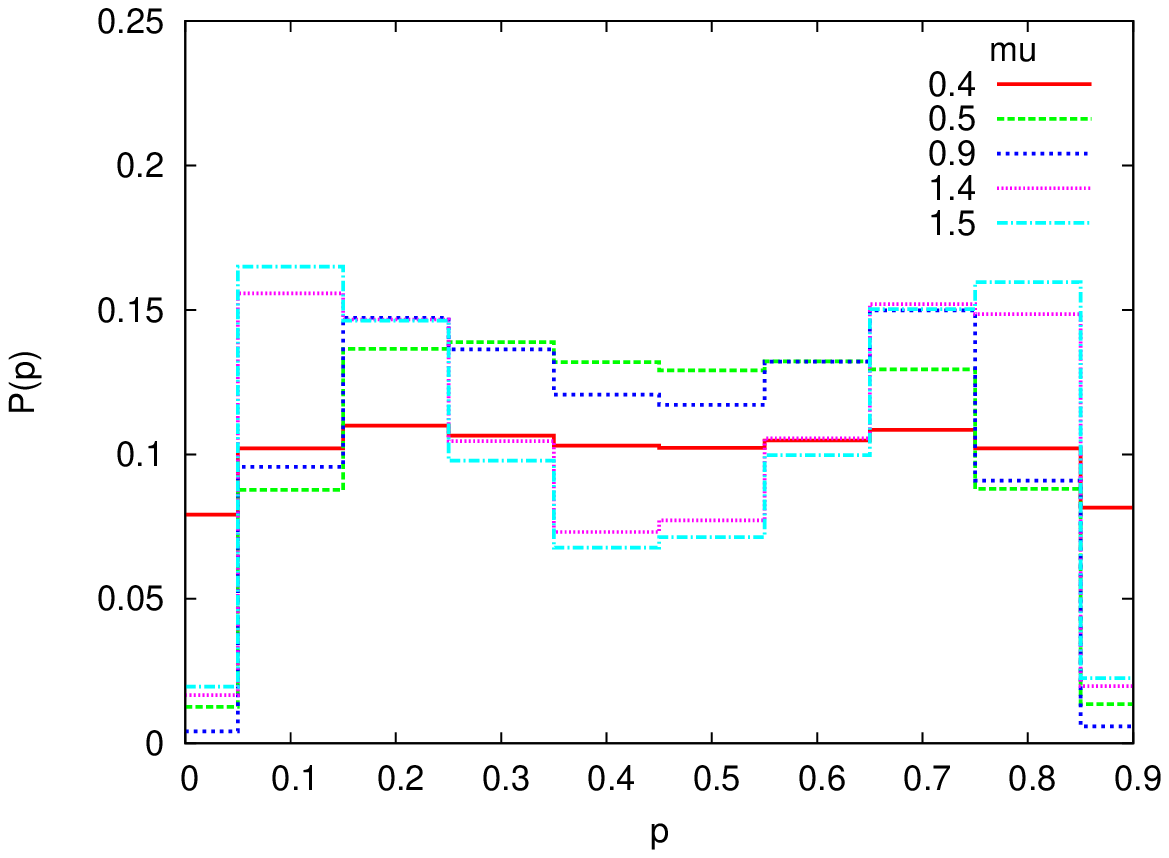}
\caption{The probability distribution of $p_i$ for various capacities $\mu$ with interaction between agents, distributed in a random network.}
\label{fig-6}
\end{figure}

For the second variant of the calculations, i.e. agents at nodes of a random network, the mean degree of the network is set to be $\lambda=5$. The results are obtained for $10^3$ agents. As before, the number of messages sent by media is $100$. The results are shown in Figs. \ref{fig-5} and \ref{fig-6}. As we see, these results are similar to those from the probabilistic variant. We see that in the case of interaction (Figs. \ref{fig-4} and \ref{fig-6}) and the largest capacity $\mu$, the opinions close to $p_i=0.5$ are even less probable than for moderate capacity. This effect is partially due to correlations, what is commented below, and partially---to the lower number of messages from other agents, what causes that fluctuations are damped slower. 

\begin{figure}[ht]
\psfrag{var}{$\text{Var}(p_i)$}
\psfrag{mu}{$\mu$}
\includegraphics[width=0.48\textwidth]{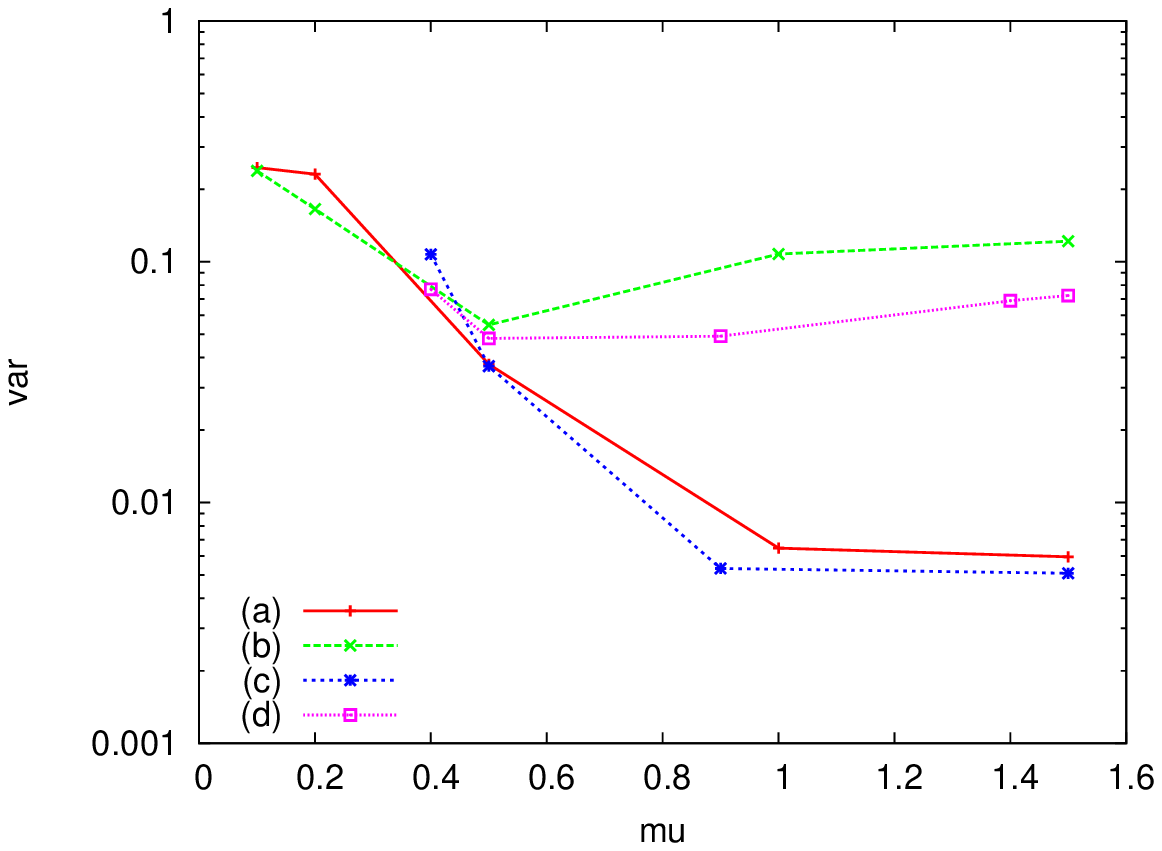}
\caption{The variance of the probability distribution $P(p_i)$ for all variants of the calculations, as dependent on the capacity $\mu$. 
Data correspond to the probabilistic variant of the calculations (a, b) and agents distributed in a random network (c, d) with (b, d) and without (a, c) interactions between agents.}
\label{fig-7}
\end{figure}

In Fig. \ref{fig-7} we show the variance of the distribution $P(p_i)$ for the probabilistic scheme and for the network, without interactions or with interactions between agents. As all obtained curves $P(p_i)$ are close to symmetric with the mean close to 0.5, a small value of variance means that
extreme opinions are absent. As we see, this is achieved for large $\mu$ without interactions. In the case with interactions, the decrease of the 
variance with $\mu$ is stopped for $\mu$ close to 0.5. This means, that the interactions prevent the agents from taking advantage of their large capacity $\mu$. The plots for the probabilistic case and for the network are quantitatively the same. In the latter case, the plots start from $\mu$=0.4, because 
the system on a network evolves very slowly when $\mu$ is small.

\begin{figure*}
\psfrag{M}{$M=$}
\psfrag{pi}{$p_i$}
\psfrag{P(pi)}{$P(p_i)$}
\includegraphics[width=0.48\textwidth]{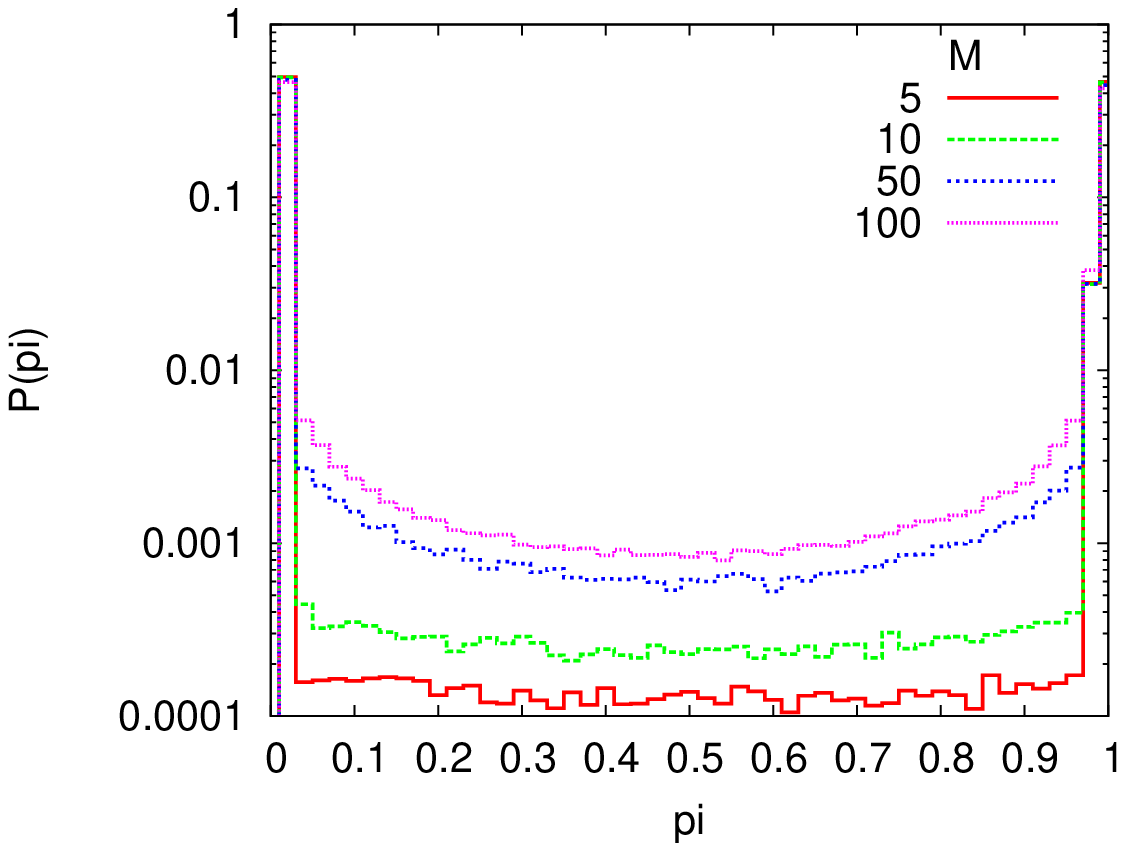}
\includegraphics[width=0.48\textwidth]{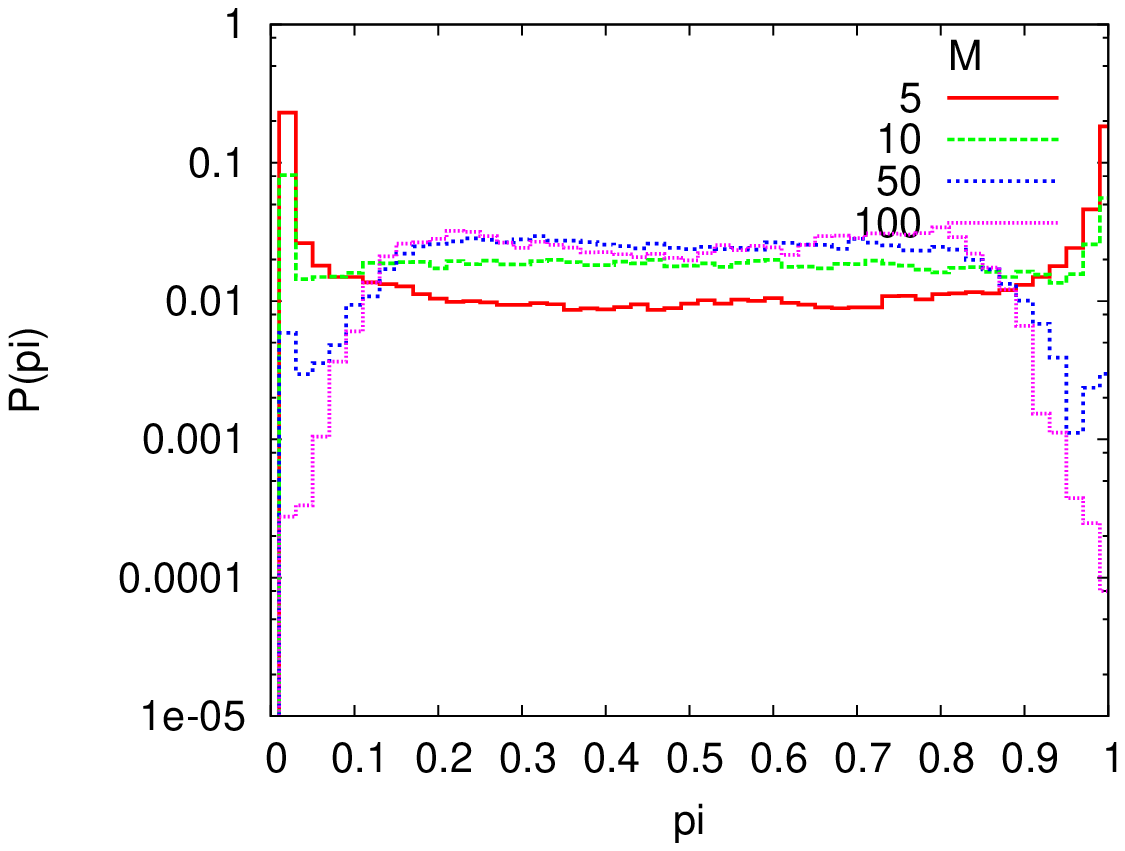}\\
\includegraphics[width=0.48\textwidth]{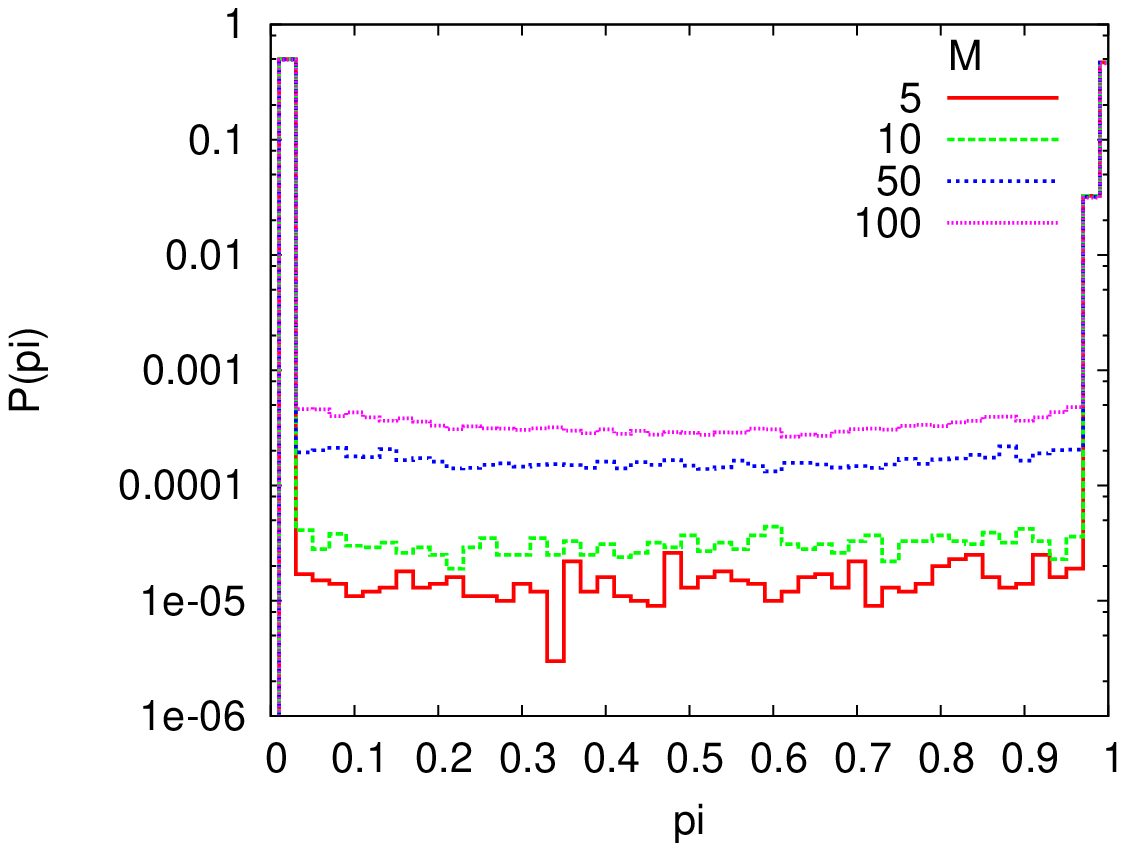}
\includegraphics[width=0.48\textwidth]{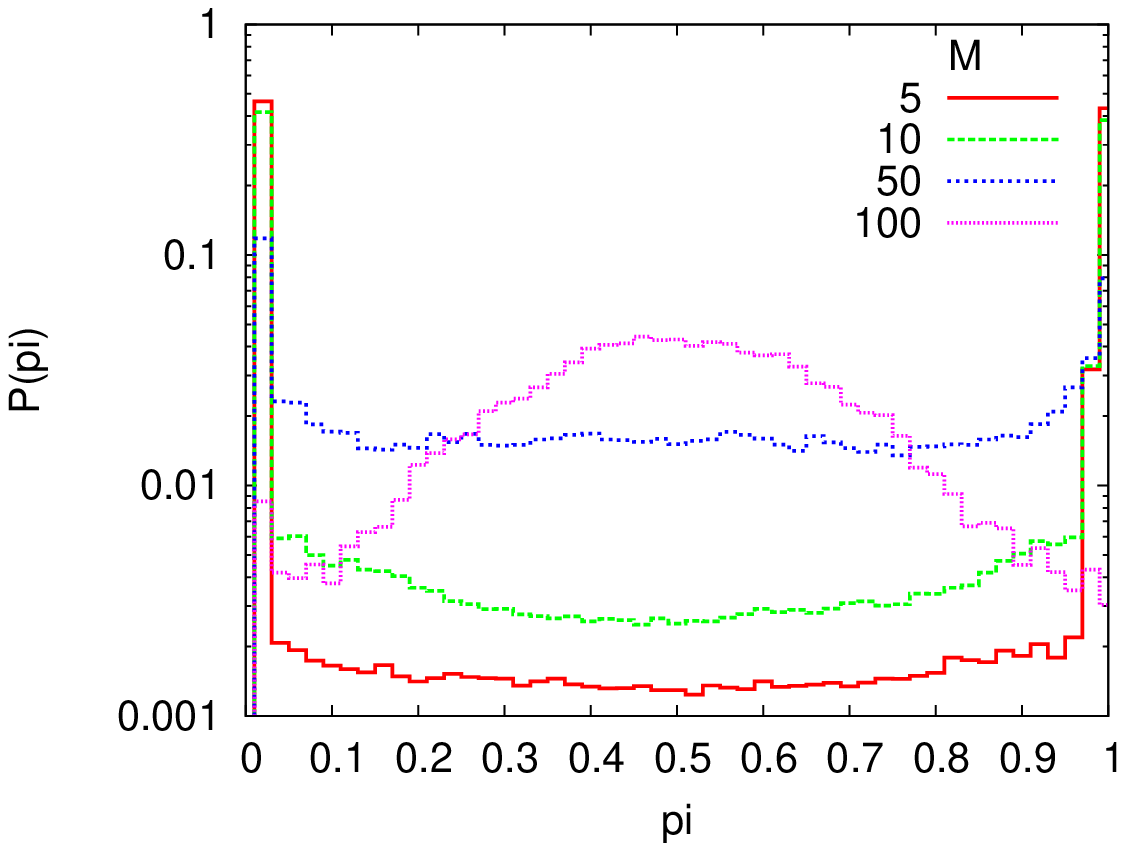}
\caption{The time evolution of the probability distribution $P(p_i)$ for various number of messageges $M$ sent to $N=10^3$ agents.
Subsequent rows correspond to model with and without interaction among agents.
Left column corresponds to $\mu=0.1$ while right is for parameter $\mu$ equal to 0.5.}
\end{figure*}

\section{Discussion}
The model presented here deals with the problem formulated by Zaller \cite{zal}: how the information provided by media is proceeded in our minds?
Although our formulation \cite{ja} is far from the original version \cite{zal}, the subject is not changed. From the Zaller's approach, we adopt the picture of agents exposed to a stream of messages. We adopt also the condition of the receipt of a message, formulated in the spirit of Deffuant model \cite{def}, i.e. with the criterion of bounded confidence. The action of this criterion depends in the model on the actual state of consciousness of a given agent, encoded by the received messages. We have then the memory effect in the sense that the agents' previous experience influences their behaviour. Further, akin to the approach of Martins \cite{mar} and in accordance with \cite{zal}, the agent's ability to receive new messages increases with his previous experience. Finally, we adopted the Zaller's postulate that opinions are constructed using the ideas most salient for a given agent. Summarizing, the model captures the time evolution of the social system, driven to large extent by the information from media, but also able to some autonomous behaviour due to the interaction between agents.

Most important result of \cite{ja} was, that agents with small mental capacity are more prone to extreme opinions. This result is reproduced here.
The main aim of this paper is to reveal the role of exchange of information between agents. We have shown that this interaction is not meaningful if the agents have small capacity, as they are in most cases unable to receive the messages. However, if their capacity is larger, the exchange of information between agents leads to the unification of their opinions. As a consequence, their opinions on the considered issues are well established.

Some aspects of this computational result are natural, but some other are counter-intuitive. It is not a surprise that exchange of information leads to a unification of opinions. The same final state was obtained in the Sznajd model \cite{sznajd} and in the Deffuant model \cite{def}; in the latter case, the condition of consensus was that the threshold should be large enough. In our model with the interaction the consensus appears always, if the number of messages is sufficiently large; just the distribution of $p_i$'s gets more and more narrow. This means that the lack of consensus is a transient effect. What is counter-intuitive is that the mean value of $p_i$, equal to 0.5 by symmetry in the absence of interaction, is different from 0.5 when the interaction is present.

In the model presented here we evade the discussion of infinitely long time, infinite number of agents and infinite number of messages, what is a commonly desired target in statistical mechanics. Instead, we concentrate on transient effects; indeed, society never attains equilibrium. We discussed two limit cases: the case when the exchange information between agents is absent, and another case when it is so dense that each agent gets five or ten times more messages from other agents, than from outside. In the second case, the stream
of informations obtained by each agent is dominated by the messages from other agents. On the contrary to the first case, these messages 
are correlated; the correlation is stronger in the case of the network, as only small number ($\lambda =5$ in the average) of agents send messsages to each agent. From the mathematical point of view, the origin of the extreme opinions is the random walk of agents in the space of issues. The goal of the model is the comparison of two realities, the sociological and the mathematical one. Our thesis is that being exposed to a stream of accidental messages has some similarities to performing a random walk. Then, the first reality can be better understood by discussing the second one.

If this is true, two conclusions are justified, both dealing with translations of mathematical facts to social reality. First mathematical fact is that the path of correlated random walk (small $\mu$, no interaction) ends more far from the initial position, than when there is no correlation (large $\mu$, no interaction). This, translated to the problem of opinions, means that slow understanding leads to more extreme opinions and less doubts. Second mathematical fact is that an attraction between trajectories of random walkers (large $\mu$, interaction) makes their spatial distribution very narrow,  with the mean of this distribution still at random position. This, translated to the problem of opinions, means that strong consensus about some issue in a finite community is always biased with respect to the accessible information on this issue.

A class of obvious and important facts remain out of scope of our model. More than often, accessible informations are so sophisticated that they remain ignored even by best experts. We have no criterion to evaluate the amount of this kind of information. Further, information from media is never complete and always biased. Again, we have no criterion to measure this bias. Actually the information from media is largely what the audience want to read; the media act then as a generalized demagogue, and the system ``media + audience'' gets an autonomy as a whole. Media, however, can hardly be described with any kind of statistics. The problem with the bias of media could be cured if the coordination center is chosen arbitrarily, as absolute true. Such a decision needs an opinion about values. Second class contains facts which are not considered in the model, but it is possible to include them.  One example is the probability distribution of the capacity $\mu$ in a given population. Further, more endowed agents could have more or less opportunities to send messages, or this opportunity could be conditioned by some bias of the message content. More generally, the idea of bounded confidence can be applied to sociological problems of communication, where the message receipt depends on the social status of the sender and of the receiver and on the state of the social bond between them \cite{scheff}. Finally---and this is third class of facts---the model allows to predict some of them. In our opinion, the result that minds more sharp are less prone to extreme opinions does belong to this category. We note also a recent critique of the Zaller model \cite{blais}, where the original version of the model \cite{zal} was confronted with the results of the 1988 Canadian elections. The authors point out that according to the statistical data, the most aware persons do not form their opinions on the basis of their predispositions. This conclusion of \cite{blais} directly agrees with our main result: most aware persons have no predispositions, if only the mutual exchange of opinions does not repress their mental independence. This indicates, that the critics of \cite{blais} does not apply to our formulation of the Zaller model.

\begin{acknowledgments}
This work was partially supported from the FP7 project SOCIONICAL, No. 231288.
\end{acknowledgments}
 

\end{document}